\input{aipcheck}

\documentclass[
    ,final            
  ]
  {aipproc}

\layoutstyle{8x11single}

\newcommand{\swf}{{\it SWIFT}}

\begin{document}

\title{Spectral curvature behavior during X-ray flares in GRB afterglow emission}


\author{F. Massaro}{
  address={Harvard - Smithsonian Astrophysical Observatory, Center for Astrophysics, 60 Garden Street, Cambridge, 02138, MA, USA}
}

\author{J. E. Grindlay}{
  address={Harvard - Smithsonian Astrophysical Observatory, Center for Astrophysics, 60 Garden Street, Cambridge, 02138, MA, USA}
}

\begin{abstract}
One of the most impressive recent discovery of \swf~ is the evidence that X-ray flares occurring during the
GRB afterglows are quite common, being observed in roughly 50\% of the afterglows.
These X-ray flares range fluences comparable with the GRB prompt emission and could be also repetitive.
Several pictures have been proposed on their origin and among them the most accepted regards the internal shock scenario,
interpreting the X-ray flares as late time activity of the GRB central engine.
We propose to describe the spectral shape of the X-ray flares adopting the same physical model recently used to interpret the GRB prompt emission:
the log-parabolic function. In particular, we show that their spectral energy distribution (SED) is remarkably curved, while no 
significant curvature appears in the underlying X-ray afterglow emission. 
In addition, the log-parabolic function is statistically favored with respect to other proposed spectral models.
By using a time resolved spectral analysis, we show the evolution of the peak energy and the curvature parameters of the SED
during the X-ray flares in two of the brightest GRBs afterglows observed by \swf.
We found that in the X-ray flares there is an anti-correlation between the peak energy and the curvature, 
as expected in a stochastic acceleration scenario.
\end{abstract}
\maketitle

\section{Introduction}
The basic scenario toward the interpretation of the X-ray flares
was described by Burrows et al. (2006, 2007), Chincarini et al. (2006, 2007) and Falcone et al. (2007).
X-ray flares in GRB early afterglows could be produced by several mechanisms,
as internal shocks, the same mechanism responsible for the GRB prompt
emission but occurring at late times or external forward shocks and external
reverse shocks. 

However, the recent \swf~ observations seem to rule out both the forward or reverse external shock scenarios. 
This picture is supported by several observational evidences, as the rapid rise and decay timescales of the X-ray flares, their multiplicity, 
the presence of an underlying continuum consistent with the same slope before and after the flare,
and the enormous increase in flux in giant X-ray flares comparable to the 
power emitted in the GRB prompt emission.

Here, we present the analysis of the time resolved spectra of the X-ray flaring activity occurring in the early afterglow
of GRB~060124 (Romano et al. 2006) observed by \swf, the birghtest ever observed by \swf~up today.
We investigate their spectral evolution in order to compare it with that observed in the GRB prompt emission,
adopting the log-parabolic function (Massaro et al. 2010).
We show that there is a characteristic behavior during the rise and the decay phases of X-ray flares.
Finally, we discuss on a the possible interpretation of this flaring activity in the framework of synchrotron emission.

For our numerical results, we use cgs units unless stated otherwise
and we assume a flat cosmology with $H_0=72$ km s$^{-1}$ Mpc$^{-1}$,
$\Omega_{M}=0.26$ and $\Omega_{\Lambda}=0.74$ (Dunkley et al. 2009).

\section{Time resolved spectral analysis}
We performed a time resolved spectral analysis of the X-ray flares in the brightest afterglow:
of GRB~060124. The time interval have been selected as shown in Figure  1.
We described the X-ray continuum (0.5 -- 10 keV) with several different spectral models:
\begin{itemize}
\item{an absorbed power-law with column density either free or fixed at the Galactic value;} 
\item{a power-law with an exponential cutoff:} 
\item{a broken power-law}
\item{the canonical GRB Band function}
\item{a blackbody model plus power-law function}
\item{a log-parabolic (i.e. log-normal) model}
\end{itemize}
All models include a Galactic column density fixed to the values reported in
the LAB survey (Kalberla et al. 2005).
\begin{figure}
\includegraphics[width=1.\textwidth]{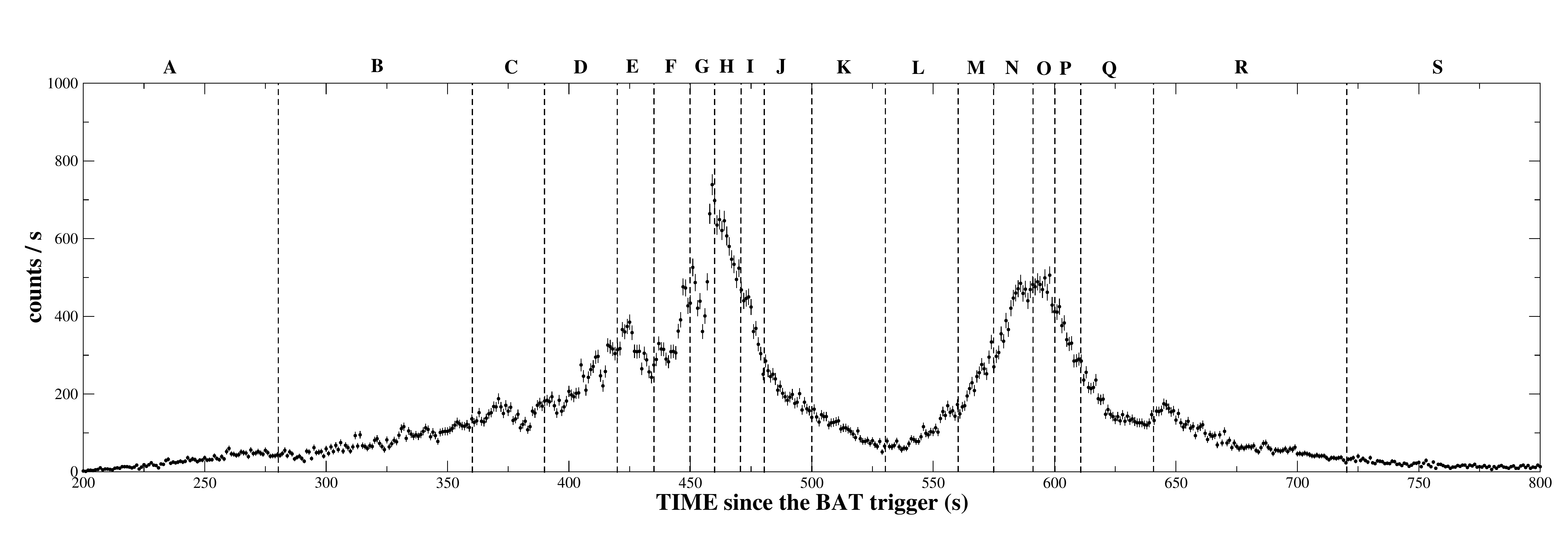}
\caption{The lightcurve of the afterglow of GRB 060124 together with the selected time intervals.}
\end{figure}

{\bf The spectra of the X-ray flares appear to be featureless and curved over a broad energy range. 
All the models adopted to describe the X-ray spectra provide unacceptable values of $\chi^2_r$ (i.e. $\geq$ 1.6)
with the only exception of the log-parabolic one.}
The latter model has been tested under the form:
\begin{equation}
F(E) = K~\left(\frac{E}{E_0}\right)^{-a-b~log(E/E_0)}~[\frac{photons}{cm^{2}~s~keV}]
\end{equation}
 where $a$ is the photon index at the pivot energy $E_0$, 
fixed to the value of 1keV, $K$ is the normalization and $b$ the spectral curvature (see Massaro et al. 2010).
We also adopted the following representation of the log-parabolic function:
\begin{equation}
F(E) = \frac{S_p}{E^2}~10^{-b~log^2(E/E_{p})}~~[\frac{photons}{cm^{2}~s~keV}].
\end{equation}
The values of the parameters $E_p$ (the location of the SED energy peak), 
$S_p$ (the peak height), and $b$  can be estimated independently 
in the fitting procedure (Tramacere et al. 2007).

Figure 2 shows the \swf~XRT spectrum of a selected time interval of the GRB~060124 afterglow 
together with the log-parabolic best-fit model and its residuals.

To investigate the X-ray flares spectral evolution, we analyzed
long exposure observations selecting several time intervals.
We chose these time intervals in order 
to avoid averaging significant spectral variations leading to an incorrect modeling of the spectral shape, 
while still conserving a sufficient number of counts (i.e. more than $\sim$ 1000 counts) to evaluate the spectral parameters and in particular, the curvature $b$.

The log-parabolic model has been successfully adopted to describe the time integrated and the time resolved 
spectra of the GRB prompt emission (Massaro et al. 2010, Massaro\& Grindlay 2011a) as alternative 
scenario, physically motivated, to the usual Band function (Band et al. 1993).
It aries from a stochastic acceleration scenario 
when terms of the 2nd order Fermi acceleration are taken into account (Kardashev 1962).
\begin{figure}
\includegraphics[width=.45\textwidth,angle=-90]{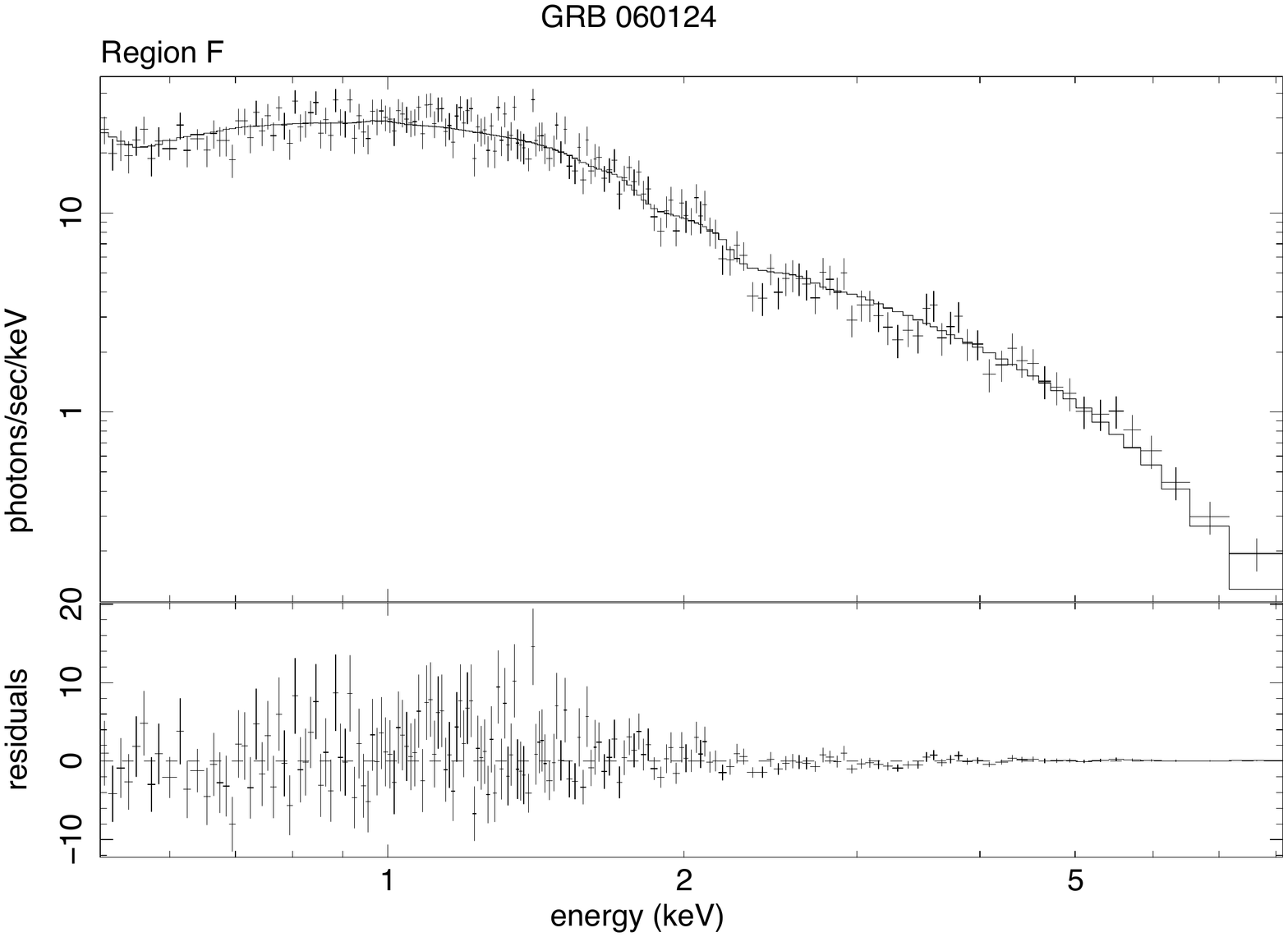}
\caption{The \swf~XRT spectrum together with the log-parabolic best-fit model and its residuals
for a selected time interval of GRB~060124 afterglow.}
\end{figure}

\section{Spectral behavior of X-ray flares}
The spectral evolution of the X-ray flares cannot described in terms of variations of the intrinsic column densities (e.g. Butler \& Kocevski 2007).
{\bf We found that the spectra of the X-ray flares are remarkably curved.
The $b$ parameter typically ranges between 0.3 and 0.8, and it is not present in the spectra of the afterglow continuum emission
underlying the X-ray flares.}
In the two GRB afterglows analyzed, we observe that the
three spectral parameters: $S_p$, $E_p$ and $b$ show a particular temporal behavior,
(Figs. 3a).
\begin{itemize}
\item{ $E_p$ is increasing when $S_p$ increases.}
\item{ $b$ decreases when $E_p$ increases.}
\item{the photon index estimated from our fitting procedure
is in agreement, within 3$\sigma$, with the expected value of the synchrotron radiation $\geq$ -2/3 
(i.e. "synchrotron line of death", e.g. Preece et al. 1998)}
\end{itemize}

\section{Stochastic acceleration}
Assuming that synchrotron losses are dominating the spectral evolution, we expect that the observed SED, 
should be narrower ($b$ increases) when $S_p$ is decreasing, but this clear is not clearly detected.

{\bf On the other hand, the most interesting observational evidence is that the curvature shows an anti-correlation
with respect to $E_p$: $b$ decreases when $E_p$ increases.}
This effect raises new clues on the spectral behavior of the X-ray flares.
It suggests that an acceleration mechanism should be always present to re-accelerate particles
hiding the radiative cooling effects, as suggested for the GRB prompt emission (Massaro \& Grindlay 2011a). 

The anti-correlation between $E_p$ and $b$ is very well known in BL Lacs objects (e.g. Tramacere et al. 2007, Massaro et al. 2008)
where it is a strong signature of stochastic acceleration occurring in their jets.
\begin{figure}
\includegraphics[width=.5\textwidth]{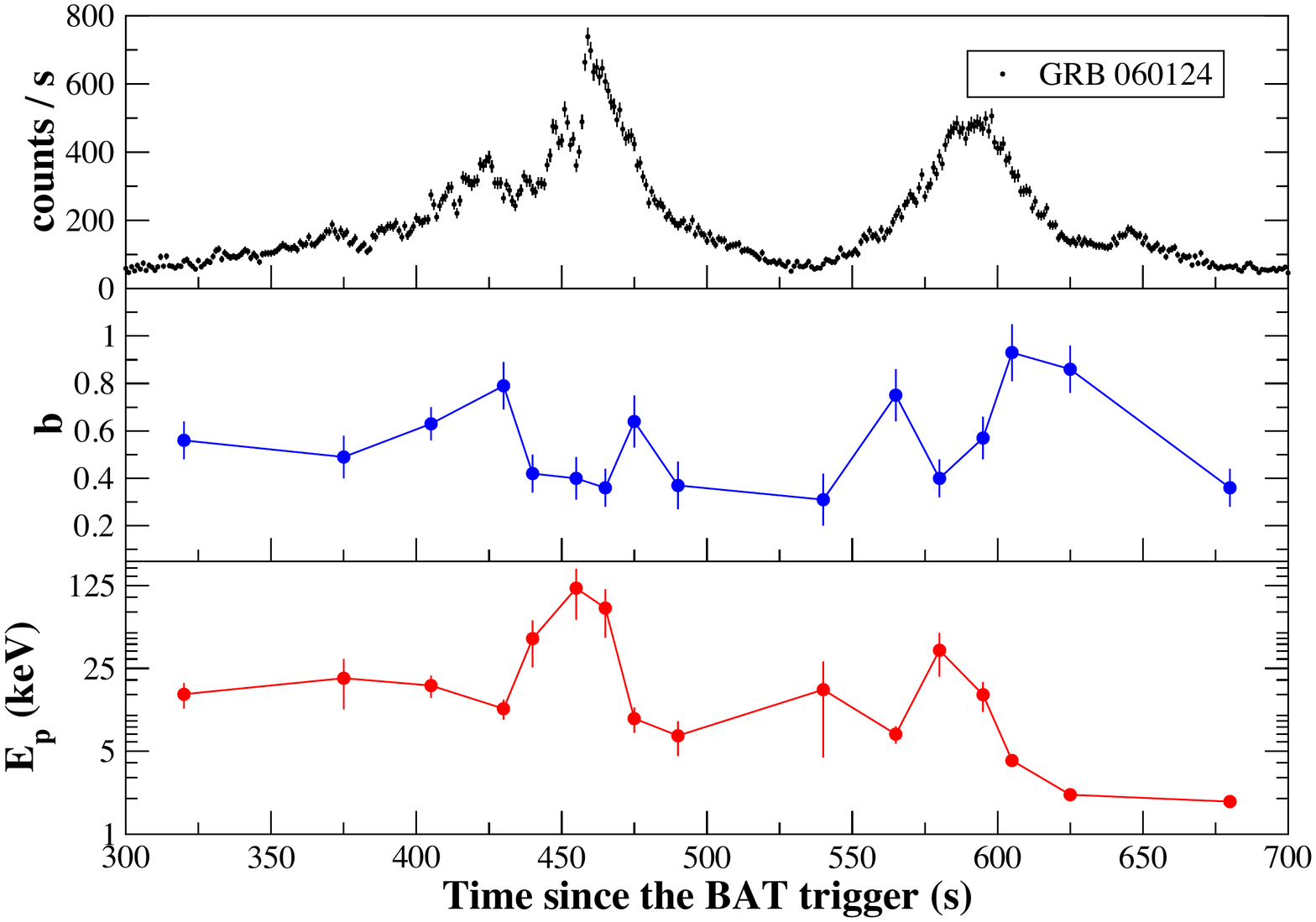}
\includegraphics[width=.5\textwidth]{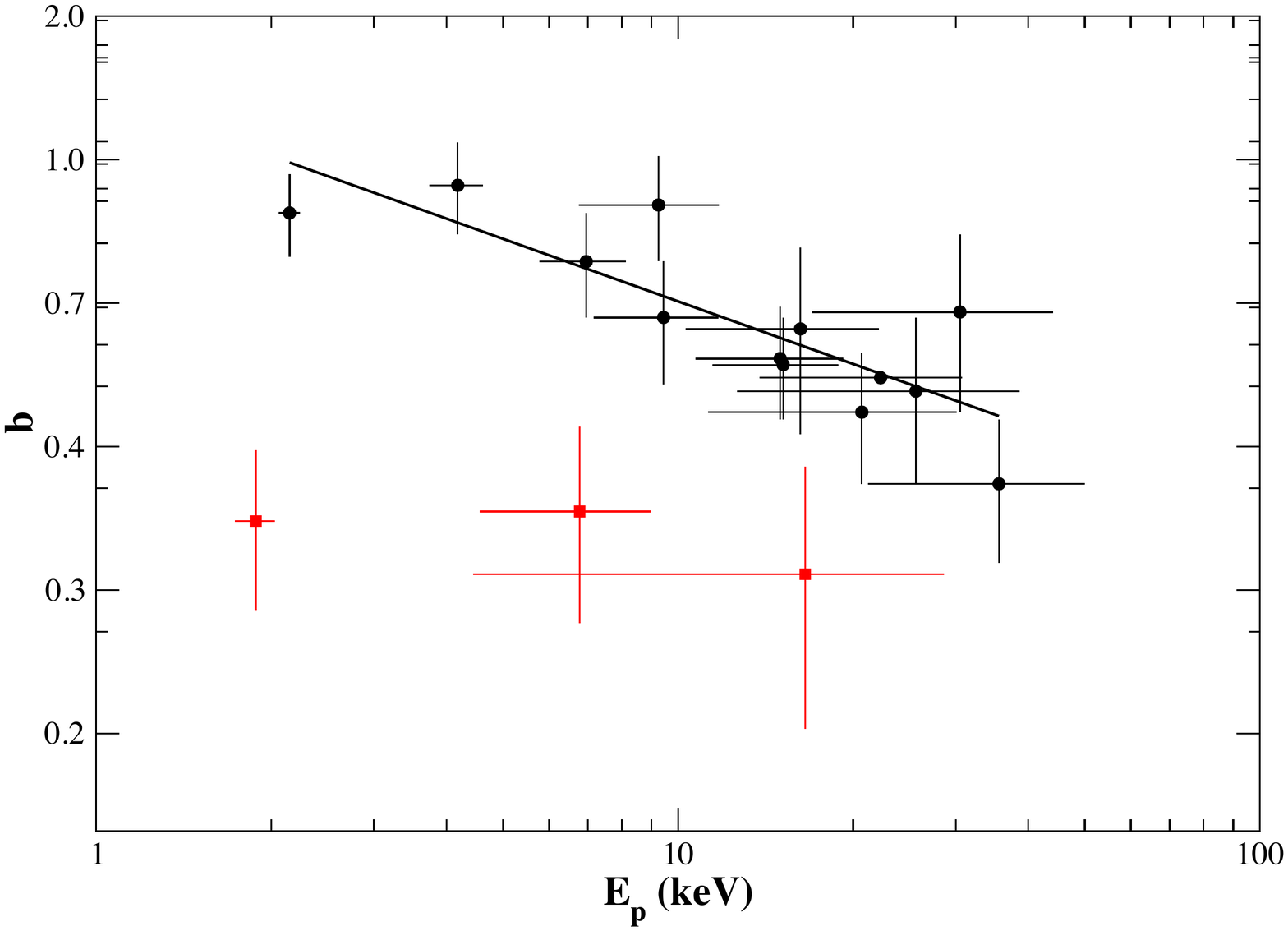}
\caption{{\it Left)} GRB~060124 afterglow light curve (upper panel). The behavior of the spectral curvature $b$ (middle panel)
in comparison with that of $E_p$ (lower panel).
{\it Right)} The correlation between $E_p$ and the curvature $b$. The red points (excluded from the statistical investigation)
refers to the time intervals J, L, S for which the flaring activity is of the order of the GRB afterglow underlying continuum.}
\end{figure}

\section{Conclusions}
The main results of our analysis are:
\begin{itemize}
\item{The anti-correlation between $E_p$ and $b$ is a strong signature of stochastic acceleration at work in X-ray flares (see Fig. 3b).}
\item{The $b$ evolution is similar with respect to that observed in GRB prompt emission,
where the spectral curvature is not drastically varying during single pulses (Massaro \& Grindlay 2011a)}
\item{We argue that {\bf synchrotron radiation} form a curved particle distribution could describe the 
X-ray flare emission during GRB afterglows (see Massaro \& Grindlay 2011b).}
\end{itemize}

\begin{theacknowledgments}
F. Massaro is grateful to R. Preece for the fruitful discussions 
and also thanks A. Cavaliere and A. Paggi for their helpful suggestions.
F. Massaro acknowledges the Foundation BLANCEFLOR Boncompagni-Ludovisi, n'ee Bildt 
for the grant awarded him in 2010 to support his research.\\
\end{theacknowledgments}

\end{document}